\documentclass[11pt,a4paper]{article}
\usepackage{epsfig,amsmath,amssymb,graphics,graphicx,color,calc,cite,psfrag,cases}

\let\id=\identity
\def\vdeg{\Lambda}
\def\fdeg{P}
\def\evdeg{\lambda}
\def\efdeg{\rho}

\def\cP{\mathcal{P}}
\def\0{{\tt 0}}
\def\1{{\tt 1}}

\begin{document}

\begin{titlepage}
\par\vskip 1cm\vskip 2em
\begin{center}

{\LARGE Group testing with Random Pools: Phase Transitions and Optimal Strategy}
\vskip 2.5em

\lineskip .5em
{\large
\begin{tabular}[t]{c}
$\mbox{M.M\'ezard}^{1,2} \phantom{m} \mbox{M.Tarzia}^{1,2} \phantom{m} \mbox{C. Toninelli}^{3,4}$
\\
\end{tabular}
\par
}
\medskip
{\small
\begin{tabular}[t]{ll}
{\bf 1} & {\it 
Universit\'e Paris-Sud, LPTMS, UMR8626, B\^at. 100, 91405
Orsay cedex, France}\\
{\bf 2} & {\it CNRS, LPTMS, UMR8626, B\^at. 100, 91405 Orsay cedex, France}\\
{\bf 3} & {\it Universit\'e Paris VI et VII, LPMA,
UMR7599, 4 Pl. Jussieu, 
Paris, 75005 France}\\
{\bf 4} & {\it CNRS,LPMA, UMR7599, 4 Pl. Jussieu, 
Paris, 75005 France}
\end{tabular}
}

\bigskip
\end{center}

\vskip 1 em

\centerline{\bf Abstract} 

\smallskip

The problem of Group Testing is to identify defective items out of a set of objects by means of  pool queries of the form ``Does the pool contain at least a defective?''. The aim is of course to perform detection with the fewest possible queries, a problem
which has relevant practical applications in different fields including molecular biology and computer science. 
Here we study GT in the probabilistic setting focusing on the regime of small defective probability and large number of objects, $p\to 0$ and $N\to \infty$. We construct
and analyze
one-stage 
algorithms for which we establish the occurrence of a non-detection/detection phase transition resulting in a sharp threshold, $\overline M$, for the number of tests.
By optimizing the pool design we construct algorithms whose detection threshold follows the optimal scaling $\overline M\propto Np|\log p|$. 
Then we consider 
two-stages 
algorithms and analyze their performance  for different choices of the first stage pools. In particular, via a proper random choice of the pools, we construct algorithms which attain the optimal value (previously determined in Ref. \cite{MezTon}) for the mean number of tests required for complete detection.
We finally discuss the optimal pool design in the case of finite $p$.

\vfill

\noindent    

\end{titlepage}

\eject

\date{\today}


\section{Introduction}

The general problem of {\sl Group Testing} (GT) is to identify 
  defective items in a set of objects. Each object  can be either {\sl defective} or OK  and we are allowed only to test groups of items via the query
``Does the pool  contain at least one defective?''.
The aim is of course to perform detection in the most efficient way, 
namely with the  fewest possible number of tests. 

Apart from the original motivation of performing efficient mass blood
testing \cite{Dorfman}, GT has been also applied in a variety of
situations in molecular biology: blood screening for HIV tests \cite{Zenios},
screening of clone libraries \cite{clone1,clone2}, sequencing by
hybridization \cite{sequencing1,sequencing2}. Furthermore it has proved
relevant for fields other than biology including quality control in product
testing \cite{control}, searching files in storage systems \cite{searching},
data compression \cite{compression} and more recently in the context of data
gathering in sensor networks \cite{sensor}. We refer to
\cite{book,review1} for reviews on the different applications of GT.

The more abstract setting of GT is the following. We have $N$ items and each one is associated with a binary random variable $x$ which takes value $1$ or $0$. We want to detect the value of all variables by performing  tests on pools of variables.  Each test corresponds to an OR function among the variables of the group,  i.e. it returns a binary variable which equals 1 (respectively 0) if at
least one variable of the pool equals 1 (respectively if all variables are 0).
Here we will only deal with this  (very much studied) choice for the tests, often referred to as the {\sl gold-standard} case. It is however important to keep in mind for future work that in many biological applications one should include the possibility of faulty OR tests \cite{Zenios,Rish}.

In all our study we will focus on
{\sl probabilistic GT} in the {\sl Bernoulli  p-scheme}, i.e. the situation in which the status of the items are i.i.d. random variables which take value one with probability $p$ and zero with probability $1-p$. 
In particular, we will be  interested in constructing efficient detection algorithms for this GT problem in the limit of large number of objects and small defective probability, $N\to\infty$ and $p\to 0$. 

In order to summarize our results we need first to introduce some terminology. The construction of any algorithm for GT involves two ingredients: 
the {\sl pool design} (the choice of the groups over which tests are performed)
and the {\sl inference procedure} (how to detect the value of the items given the result of the tests). The pool design can be composed by one or more stages of
parallel queries.
For
{\sl one-stage} (or {\sl fully non-adaptive}) algorithms 
 all tests are specified in advance: the choice of
the pools  does not depend on the outcome of the
tests. This would be in principle the easiest procedure for several
 biological applications. Indeed the test procedure 
can be destructive for the objects and repeated tests on the same
sample require more sophisticated techniques.  However
the number of tests required by
fully non-adaptive algorithms can be 
much larger than for adaptive ones. The
best compromise for most screening procedures \cite{Knill}
is therefore to consider {\sl two-stage} algorithms 
with a first stage containing a set of
predetermined pools (tested in parallel) and a second stage whose
pools are chosen depending on the outcomes of the first
stage, i.e. after an inference procedure which uses the results of the first stage. Concerning the inference procedure,
there exist both {\sl exact} and {\sl approximate
algorithms}
which lead after the last stage to 
detect the value of all variables with certainty or with high probability, respectively.

Here we will construct one-stage approximate algorithms and two-stage exact algorithms. In both cases  the pool design for the first stage will involve random pools and we will focus on the case $N\to\infty$ and $p\to 0$ with $p=1/N^{\beta}$ (the case $\beta=0$ stands for $p\to 0$ after $N\to\infty$). This choice was first discussed by Berger and Levenshtein in  the two-stage setting in \cite{Berger} where they proved that for $\beta\in(0,1)$  the minimal number of tests  optimized over all exact two-stage procedure, $\overline T(N,p)$ is proportional to $ Np|\log p|$.

In the one-stage case we will establish the occurrence of a {\sl phase transition}: considering two simple inference algorithms, we identify  a threshold $\overline M$  such that the probability of making at least one mistake in the detection goes to one (respectively to zero) when $N\to\infty$ if the number of tests $M$ is below (respectively above) $\overline M$. By optimizing over the pool distribution, we will  construct algorithms for which the detection threshold shows the optimal scaling  $\overline M=(1-\beta)(\beta)^{-1}(\log 2)^{-2} Np|\log p|$.

Recently in Ref.~\cite{MezTon} the value of the prefactor of $\overline T$ has been
determined exactly when $\beta\in[0,1/2)$ for two-stage procedures. More precisely, the
authors have shown that:
$\lim_{N\to\infty}\overline T/(Np|\log p|)=1/(\log 2)^2$.
Here we will  discuss the performance of two-stage algorithms for different choices of the first stage pool design. In particular we will show that the optimal value is obtained on random pools with a properly chosen fixed number of tests per variable and of variables per test (regular-regular case) and also when the number of tests per variable is fixed but the number of variables per test is Poisson distributed (regular-Poisson case). On the other hand we will show that this optimal value can never be attained  in the Poisson-regular or in the Poisson-Poisson case.
Finally, we discuss the optimal pool design when $N\to\infty$ and $p$ is held fixed. 

The paper is organized as follows: In Sec.~\ref{factorgraph} we introduce the factor graph representation of the problem in the most general case. In Sec. \ref{inference} we describe the first simple  inference procedure which allows to identify the {\emph {sure variables}}. In Sec.~\ref{one} we analyze one-stage approximate algorithms, while in Sec. \ref{two} we turn to the two-stage exact setting. Finally, in Sec. \ref{conclusions} we give a perspective of our work in view of applications.

\section{Pool design: factor graph representation and random pools}
\label{factorgraph}

\begin{figure}
\begin{center} 
\psfrag{m}[][]{{\large{m}}}
\psfrag{l}[][]{{\large{l}}}
\psfrag{n}[][]{{\large{n}}}
\psfrag{i}[][]{{\large{i}}}
\psfrag{a}[][]{{\large{a}}}
\psfrag{b}[][]{{\large{b}}}
\psfrag{c}[][]{{\large{c}}}
\psfrag{d}[][]{{\large{d}}}
\psfrag{j}[][]{{\large{j}}}
\psfrag{k}[][]{{\large{k}}}
\includegraphics[scale=0.45]{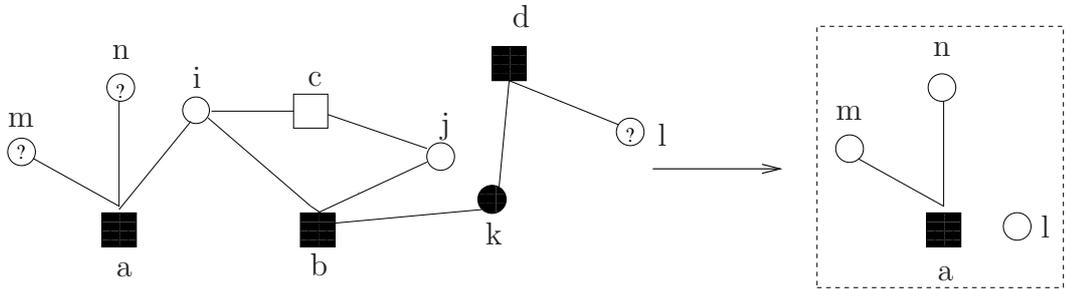}
\end{center}
\caption{Left: The bipartite graph corresponding to a single stage.
Circle (squares) represent variable (test) nodes. We depict
by filled (empty) squares the tests with outcome one (zero, respectively).
Variables $i$ and $j$ are sure zeros, variable $k$ is a sure one, variable $m$,$n$ and $l$ are undetermined. Right: The corresponding reduced graph where the sure variables ($i,j,k$) and the strippable tests ($b,c,d$) have been erased. Note that one of the three variables, $l$, is isolated.}
\label{graph}
\end{figure}

As we have explained, a GT algorithm can involve one or more stages of parallel tests. 
The best way to define  the pool design of each stage is in term of a factor graph representation. 
We build a graph with two types of vertexes: each variable is a vertex (variable node) and each test is also a vertex 
(function node). Variable (function) nodes will be
 denoted by indexes $i,j,\dots$ ($a,b,\dots$) and depicted by a circle (square).
Whenever a variable $i$ belongs to test $a$ 
we set an edge between vertex $i$ and $a$.
Thus if $N$ is the overall number of items and $M$ the number of parallel tests in the stage, we obtain a {\sl bipartite graph} with $N$ variable nodes and $M$ test nodes with edges between variables and tests only (in Fig \ref{graph} we depict a case with $N=6, M=4$). 
 
We will denote  by $\vdeg_{i}$ (by $\fdeg_{i}$) 
the fraction of variable nodes (function nodes) of degree $i$
and use a practical representation of these {\sl degree profiles}, standard in coding theory, in terms of  their generating functions 
$\vdeg(x) = \sum_{n\ge 0}\vdeg_n\, x^n$ and $\fdeg(x) = \sum_{n\ge 0}
\fdeg_n \, x^n$. 
The average variable node (resp. function node) degree is given by
$\sum_{n\ge 0 }\vdeg_n\,  n =\vdeg'(1)$ 
(resp. $\sum_{n\ge 0 }\fdeg_n\,  n =\fdeg'(1)$).

Both in the one and two stage case we will use pool designs with a first stage based on a randomly generated factor graph. We will consider different 
possible distributions, but in all cases they will be uniform
over the set of graphs for a fixed choice of the
degree profiles $\Lambda(x)$ and $P(x)$. Thus
the probability $\evdeg_{l}$ (respectively $\efdeg_k$)  that a randomly chosen 
edge in the graph is adjacent to a variable node (resp. function node) of
degree $l$ (degree $k$) are given by
\begin{eqnarray}
\evdeg_{l} = \frac{l\vdeg_l}{\sum_{l'}l'\vdeg_{l'}}\, ,\;\;\;\;\;\;\;\;\;\;
\efdeg_{k} = \frac{k\fdeg_k}{\sum_{k'}k'\fdeg_{k'}}\, ,
\end{eqnarray}
which are derived by noticing that the graph $F$ contains 
$nl\vdeg_l$ (resp. $mk\fdeg_k$) edges adjacent to variable nodes of degree $l$
(resp. function nodes of degree $k$). 
We also define the {\sl edge perspective degree profiles} as
$\evdeg[x] \equiv \vdeg'[x]/\vdeg'(1)$ and $\efdeg[x]\equiv
\fdeg'[x]/\fdeg'(1)$, namely
\begin{eqnarray}
\evdeg[x] = \sum_{l=1}^{l_{\rm max}}\evdeg_l\, x^{l-1}\, ,\;\;\;\;\;\;\;\;\;\;
\efdeg[x] = \sum_{k=1}^{k_{\rm max}}\efdeg_k\, x^{k-1}\,.
\end{eqnarray}
Note that the number of checks $M$ can also be written in terms of these sequences,
because the mean degree of variables, $\langle l \rangle=\vdeg'(1)$, and the
mean degree of tests, $\langle k \rangle= \fdeg'(1)$, are related by $N\langle
l \rangle= M \langle k \rangle$. As $\vdeg'(x)=\vdeg'(1) \evdeg(x)$ and
$\vdeg(1)=1$ we get $\langle l \rangle =1/\int \evdeg(x) dx$. Therefore 
\begin{equation}
\label{M}
M=N
\frac{\int \efdeg(x)dx}{\int \evdeg(x) dx}.
\end{equation}

\section{First inference step: sure and isolated variables}
\label{inference}

After the first stage of tests we will either use an inference
procedure to identify the result (in the one stage case) or choose a
new set of pools based on the outcomes of the previous tests (in the two stage case).
In our problem, the prior distribution of the $N$ variables, $x=(x_1,\dots,x_N)$, is Bernoulli: $B_p(x)=\prod_{i=1}^N p^{x_i}(1-p)^{1-x_i}$. Given the outputs of the tests, the
inference problem consists in finding  the configuration $\bar x$ which maximizes
\begin{equation}
\label{infe}
P(x)=\frac{B_p(x)}{Z}\prod_{a=1}^M\id(T_a(x)=t_a)\ .
\end{equation}
Here $t_a$ is the value of test $a$ and $T_a(x)=0$ if $\sum_{j\in {\cal{N}}_a}
x_j>0$, $T_a(x)=0$ otherwise, where ${\cal{N}}_a$ is the pool of variables connected
to $a$. 

Since the minimization of the above function is in general a very
difficult task, we start by checking whether some variables are identified
with certainty by the first stage and then try to extract information on the
remaining variables (see Fig.\ref{graph}). The first observation is that in
order for a variable to be a {\sl sure zero} it should belong to at least one
test with outcome zero. On the other hand in order to be a {\sl sure one} it
should belong to at least one positive test in which all the other variables are sure
zeros. Variables that are neither sure zeros nor sure ones are the {\sl
  undetermined variables}.

We start by noticing that if a test contains only zeros, or if it contains at
least a sure one, then it does not carry any information on the undetermined
variables. We call such a test {\sl strippable}, as is the case for tests
$b,c,d$ in Fig.\ref{graph}. It is then immediate to verify that we have no
information on a variable if it is undetermined and all the tests to which it
belongs are strippable. We call such undetermined variable {\sl isolated}, as
is the case for variable $l$ in Fig.\ref{graph}. The above terminology is
motivated by the fact that all the information on the undetermined variables
is encoded in a {\sl reduced graph} (see right part of Fig. \ref{graph})
constructed via the following stripping procedure: erase all variable nodes
which correspond to sure variables and all test nodes which are strippable
(note that isolated variables are those that are not connected to any test in
the reduced graph). Therefore the inference problem corresponding to the
minimization of \eqref{infe} can be rephrased as a Hitting Set problem on the
corresponding reduced graph \cite{MT}.

Given a variable $i$ and a choice of the pools, the probability $p_{s0}^i$
(respectively $p_{s1}^i$) that $x_i$ is a sure zero (resp. a sure one) can be
found as follows. Let us denote by ${\cal{N}}_a$ (${\cal{N}}_i$) the set of
variable (resp. the set of tests) nodes connected to test $a$ (resp. variable $i$).
We introduce the indicator $G_i( x)$ that $x_i$ is a sure $0$ as well as  the indicator $V_{i}( x)$ that $x_i$ is a sure one:
\begin{equation}
G_i( x)=(1-x_{i})\left\{1-\prod_{a\in {\cal{N}}_i}W_{i,a}( x)\right\}\ ,
\end{equation}
\begin{equation}
\label{Vj}
V_{i}( x)=x_i\left\{1-\prod_{a\in {\cal{N}}_i}\left[1-\prod_{\stackrel{k\in {\cal{N}}_a}{k\neq i}} 
G_k( x)\right]\right\}.
\end{equation}
which are expressed in terms of 
\begin{equation}
\label{wia}
W_{i,a}( x)=
1-\prod_{\stackrel{j\in {\cal{N}}_a}{j\neq i}}(1-x_j)
\end{equation}
Then  $p_{s0}^i$ and  $p_{s1}^i$ are given by:
\begin{equation}
p_{s0}^i:=\sum_{ x} B_p(x) G_i( x)
\label{ps0}
\end{equation}
\begin{equation}
p_{s1}^i:=\sum_{ x} B_p(x) V_i( x)
\label{ps1}
\end{equation}
where the sum is over all $ x\in\{0,1\}^N$.

It is clear that Eq. \eqref{ps0} for $p_{s0}^i$ involves
only the variables at graph distance two from $i$. Thus, if $i$ does not
belong to a loop of length four in the factor graph, $W_{i,a}$ are independent
variables and the mean over the variable values in \eqref{ps0} can be easy
carried out yielding
\begin{equation}
\label{ps0noloops}
p_{s0}^i=(1-p) \left[ 1-\prod_{a\in {\cal{N}}_i}\left(1-(1-p)^{k_a-1}\right) \right]
\end{equation}
where $k_a=|{\cal{N}}_a|$ is the number of variables which belong to test $a$.
Then, for any given choice $\Lambda,P$ of the degree profiles of the random
factor graph, if the probability that two tests have degree $k$ and $k'$
factorize, we can easily perform the mean over the uniform distribution for
the factor graphs. This leads to a value which does not depend anymore on the
index $i$ and can be rewritten as $p_{s0}^i=(1-p)S_0$ with

\begin{equation}
\label{S0}
S_{0}:=\sum_{l}\Lambda_l \left(1-\left(1-\sum_k\rho_k(1-p)^{k-1}\right)^{l}\right)=1-\Lambda[1-\rho[1-p]].
\end{equation}

Formula \eqref{ps1} for $p_{s1}^i$ involves only variables at distance at most
$4$ from $i$. If the ball centered in $i$ of radius 4 does not contain any
loop,
we can perform easily the mean over the variables in \eqref{ps1} and get 
$p_{s1}^i=pS_1$ with
\begin{equation}
\label{S1}
S_1:=1-\Lambda\left[1-\rho\left[(1-p)(1-\lambda[1-\rho[1-p]]\right]\right].
\end{equation}

The probability that $t_a$ is strippable ($R^a$), $x_i$ is an isolated zero
($I_0^i$) and $x_i$ is an isolated one ($I_1^i$) are instead given by

\begin{equation}
\label{reducible}
R^a=\sum_{ x}B_p(x)\left[\prod_{j\in {\cal{N}}_a}^N(1-x_j)+1-\prod_{j\in {\cal{N}}_a}^N (1-V_j( x))\right]
\end{equation}
\begin{equation}
\label{fine1}
I_0^i=\sum_{ x}B_p(x)(1-x_i)\prod_{a\in {\cal{N}}_i}\left(1-\prod_{\stackrel{j\in {\cal{N}}_a}{j\neq i}}(1- V_j(x))\right)
\end{equation}

\begin{equation}
\label{fine2}
I_1^i=\sum_{ x}B_p(x) x_i\prod_{a\in {\cal{N}}_i}\left(1-\prod_{\stackrel{j\in {\cal{N}}_a}{j\neq i}}(1- V_j)\right).
\end{equation}

In this case, if there is no loop in the ball of radius 6 centered on $i$,
we can easily perform the mean over the variables and 
over the random graph distribution which yield 
$I_0=(1-p)I$ and $I_1=pI$ with
\begin{equation}
I=\Lambda\left[1-\rho[1-p\tilde S_1]\right],
\label{I}
\end{equation}
with
\begin{equation}
\label{S1tilde}
\tilde S_1:=1-\lambda\left[1-\rho\left[(1-p)(1-\lambda[1-\rho[1-p]]\right]\right].
\end{equation}

\section{One-stage algorithms}
\label{one}

In this section we analyze one-stage algorithms when the number of items, $N$, goes to infinity and the defect probability, $p$, goes to zero as $p=1/N^{\beta}$ with $\beta>0$. 
When constructing the pools we  use random graph ensembles of two types: either  regular-regular (R-R) graphs (fixed connectivity both for test and variable nodes) or regular-Poisson (R-P) graphs (fixed connectivity for variables, Poisson distribution for the test degree).
 As for the inference procedure we will consider two types of algorithms: Easy Algorithm (EA) and Belief Propagation (BP). We will show that both  undergo a non-detection/detection phase transition when one varies the number of tests, $M$: we identify a threshold $\overline M$ such that for $M<\overline M$  the overall detection error goes (as $N\to\infty$) to one while for $M> \overline M$ it goes to zero. When $\beta<1/3$ 
we can establish analytically the value of $\overline M$ which turns out to be equal for the two algorithms:
EA and BP have  the same performance in the large $N$ limit. We will explain why  this transition is robust and we will optimize the pool design (i.e. choice of the parameters of the regular-regular and regular-Poisson graphs) to obtain the smallest possible $\overline M$. The resulting algorithms have a threshold value which satisfies $\lim_{N\to\infty}\overline M/(Np|\log p|)=(1-\beta)\beta^{-1}(\log 2)^{-2}$. This is the same scaling in $N$ and $p$ as for the optimal number of tests in an exact two-stage algorithm, albeit with a different prefactor.

\subsection{Pool design}
\label{poolone}

 Given a random graph ensemble,  we denote by $M$ the number of test nodes, by $K$
the mean degree of tests (which also coincides with the degree of each test in the R-R case) and by $L$ the degree of each variable and we let
 \begin{equation}
\label{parameters}
M=cNp\log N,~~K=\alpha/p, ~~L=MK/N= c\alpha\log N.
 \end{equation}
The degree profile polynomials are:
$$\Lambda^{R-R}[x]=x^L,~~\lambda^{R-R}[x]=x^{L-1},
~~P^{R-R}[x]=x^K,
~~\rho^{R-R}[x]=x^{K-1}$$
$$\Lambda^{R-P}[x]=x^L,~~\lambda^{R-P}[x]=x^{L-1},~~
P^{R-P}[x]=
\rho^{R-P}[x]=e^{K(x-1)}.$$
Then, if the hypotheses on the absence of short loops which lead to (\ref{S0}), (\ref{S1}) and (\ref{I}) are valid,
 the probabilities $S_0$, $S_1$ and $I$ are given  in the R-R case by:
\begin{equation}
S_0=1-(1-(1-p)^{K-1})^L,
\label{S0rr}
\end{equation}
\begin{equation}
\label{S1rr}
S_1=1- \left \{ 1-(1-p)^{K-1}[1-(1-(1-p)^{K-1})^{L-1}]^{K-1} \right \} ^L,
\end{equation}
\begin{equation}
\label{S1rrtilde}
\tilde S_1=1-\left \{ 1-(1-p)^{K-1}[1-(1-(1-p)^{K-1})^{L-1}]^{K-1}\right \}^{L-1},
\end{equation}
\begin{equation}
\label{Irr}
I=\left(1-(1-p \tilde S^1)^{K-1}\right)^{L}.
\end{equation}
In  the R-P case they are given by:
\begin{equation}
\label{S0rp}
S_0=1-(1-\exp(-Kp))^L \ ,
\end{equation}
\begin{equation}
\label{S1rp}
S_1=1-\left(1-\exp\left(-Kp -K(1-p) (1-e^{-Kp})^{L-1}     \right)   \right)^L  \ ,
\end{equation}
\begin{equation}
\label{S1rptilde}
\tilde S_1=1-\left(1-\exp\left(-Kp -K(1-p) (1-e^{-Kp})^{L-1}     \right)   \right)^{L-1}  \ ,
\end{equation}
\begin{equation}
I=\left(1-\exp(-Kp\tilde S^1)\right)^{L}.
\label{Irp}
\end{equation}

It is easy to verify that
in leading order when $N\to\infty$ and $p\to 0$ the above quantities for the regular regular and regular Poisson case coincide. In  particular if we set $p=N^{-\beta}$ they are given by
\begin{equation}
S_0\simeq 1-N^d 
\label{S0leading}
\end{equation}
 \begin{numcases}{S_1\simeq }
\nonumber
( c\alpha \log N  ) e^{-\alpha (1+N^{d+\beta}/b)} & {\mbox{if}} $\beta+d>0$\\
 1-N^d & {\mbox{if}} $\beta+d<0$\\
 1-N^{- c\alpha | \log(1-\exp(-2\alpha))|} & {\mbox{if}} $\beta+d=0$ \nonumber
\label{S1leading}
\end{numcases}
and
 \begin{numcases}{I\simeq }
 \nonumber
(c\alpha\log N)^{c\alpha\log N} e^{-\alpha^2 c\log N(1+N^{d+\beta}/b)}
& {\mbox{if}} $\beta+d>0$\nonumber\\
N^d & {\mbox{if}} $\beta+d<0$\\
N^d & {\mbox{if}} $\beta+d=0$\nonumber
\label{Ileading}
\end{numcases}
where we set
$b=b(\alpha)=(1-\exp(-\alpha))$ and $d=d(\alpha,c)=-c\alpha|\log b|$, for $N\to\infty$. 

Let us discuss in what range of $\beta$ one expects the above asymptotic
behaviors to be valid. As explained in section \ref{inference}, the only
hypothesis in their derivation consists in neglecting the presence of some
short loops in a proper neighborhood of the chosen variable. In particular the
equation for $S_0$ is valid if we can neglect the presence of loops of length
four through a given variable. Consider for definiteness the R-R case. The
probability of having at least one loop of length four through $i$, $P(L_4)$,
verifies
$$P(L_4)\leq L^2 N \frac{\binom{M}{L-1}}{\binom{M}{L}}\simeq \frac{(\log p)^2}{N p^2}$$
which goes to zero for $\beta<1/2$. Thus we are guaranteed that \eqref{S0rr}
is correct in this regime.
By the same type of reasoning, we can show that the  formulas for $S_1$ and $I$
are valid respectively for $\beta<1/4$ and $\beta<1/6$. 
However through the following heuristic argument, one can expect that the
formula for $S_1$ (resp. $I$) be correct in the larger regimes $\beta<1/2$
(resp. $\beta<1/3$). Indeed, when we evaluate $S_1$ we need to determine
whether variables at distance $2$ from a variable $i$ are sure zeros. We
expect the probability of this joint event to be well approximated by the
product of the single event probabilities if the number of tests that a
variable at distance $2$ from $i$ shares with the others is $\ll L$ and if the
number of variables that a test at distance $3$ from $i$ shares with the
others is $\ll K$. Both conditions are satisfied if $\beta<1/2$
 (the
probability that a test at distance $3$ belongs to more than one variable
at distance $2$ goes as $(1-K/N)^{LK}$ and the probability that a variable at
distance $4$  belongs to more than one  test at distance $3$ goes as
$(1-K/N)^{L^2K}$). For $I$ the argument is similar but, since we have a
further shell in tests and variables to analyze in order to determine whether
a variable is isolated or not, we get an extra factor $KL$ in the exponents
which lead to the validity of the approximations only for $\beta<1/3$.

\subsection{Easy algorithm (EA)}
\label{EA}

A straightforward inference procedure is the one that fixes the sure variables
to their value and does not analyze the remaining information carried by the
tests, thus putting to zero all other variables (since $p<1/2$). We call this
 procedure {\sl Easy Algorithm} (EA). By definition the probability that a
variable is set to a wrong value, $E_{bit}$, is given by $E_{bit}=p-pS^1$. In
the hypothesis of independent bit errors, i.e. if we suppose that the
probability $E_{tot}$ of making at least one mistake satisfies
$E_{tot}=1-(1-E_{bit})^N$ and if $\beta<1/2$ (see the discussion at the end of
previous section), we can apply \eqref{S1leading} which yields
\begin{numcases}{E_{tot}\simeq }
\label{Etot}
 \nonumber
1-\exp(-N^{1-\beta}) & {\mbox{if}} $\beta+d>0$\\
1-\exp(-N^{1-\beta+d}) & {\mbox{if}} $\beta+d<0$\\
1-\exp(-N^{1-\beta+c\alpha\log(1-\exp(-2\alpha))} & {\mbox{if}} $\beta+d=0$\nonumber
\end{numcases}
both for the R-R and R-P graphs. Therefore EA displays a phase
transition in the large $N$ limit, when one varies the parameter $c=L/(\alpha \log N)$ from a region at $c<\bar c(\alpha)$ in which
the probability of at least one error, $E_{tot}$, goes to one, to a region $c>\bar c(\alpha)$ where it goes to
zero. The threshold of this regime
is given by 
\begin{equation}
\label{tresholdEA}\bar c(\alpha)=\frac{1-\beta}{\alpha |\log(1-\exp(-\alpha))|}\end{equation} 
The most efficient pools, within the R-R and R-P families, are obtained by
minimizing $\bar c(\alpha)$ with respect to $\alpha=K p$. The value of the
optimal threshold $\tilde c=\min_{\alpha}\bar c(\alpha)$ and the parameter
$\tilde \alpha$ at which the optimal value is attained, namely
$c(\tilde\alpha)=\tilde c$, are
 $$\tilde c=\frac{1-\beta}{(\log 2)^2}, ~~~\tilde\alpha=\log 2.$$
This, together with \eqref{parameters}, gives a threshold
\begin{equation}\overline M=Np|\log p|(1-\beta)\beta^{-1}(\log 2)^{-2}
\label{overM}
\end{equation}
for the number of tests. Note that the threshold in the case $\beta=0$, i.e.
if we send $p\to 0$ after $N\to\infty$, is infinite. This corresponds to the
fact that for any choice $M=CNp|\log p|$ and $K=\alpha/p$ the bit error
$p(1-S_1)$ stays finite when $N\to\infty$, since $K$ and $L$ depend only on
$p$.

\begin{figure}
\begin{center} 
\includegraphics[angle=0,scale=0.4]{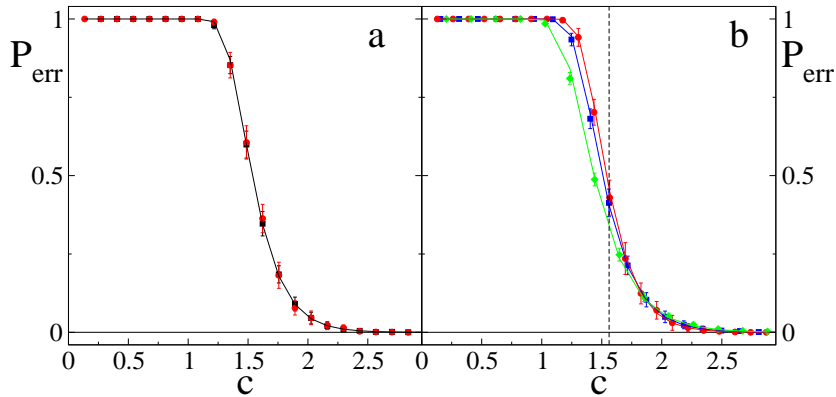}
\end{center}
\caption{{\bf a)} Error probability as a function of $c$ using EA (red circles) and BP (black squares) for a regular-regular graph. The graph  parameters are chosen as in \eqref{parameters} with $p = N^{-\beta}$, $\beta = 1/4$, $\alpha=\tilde\alpha=\log 2$ and $N = 43321$. The continuous line corresponds to the theoretical prediction of Eqs.~\eqref{Etot}.
{\bf b)}  Error probability  as a function of $c$ for EA. We set again $\beta=1/4$ and $\alpha=\log 2$, while we choose
$N = 1109$ (green diamonds), $N = 10401$ (blue squares), and $N = 63426$ (red circles). The vertical dashed
line corresponds to the threshold $\overline{c}$, given by Eq.~\eqref{tresholdEA}.}
\label{transition}
\end{figure}

In order to verify the above results and the approximations on which they are
based we have performed numerical simulations in the case of the R-R graph
with $\beta=1/4$, $\alpha=\tilde \alpha$ and different values of $c$. 
The results we obtain confirm that in this regime bit errors can be regarded as independent and  formulas \eqref{S0rr}--\eqref{Irr} are valid. 
The
values of $E_{tot}$ as a function of $c$
are depicted in Fig. \ref{transition}a for different values of $N$. 
The value of the threshold connectivity and the form of the
finite size corrections for the total error (continuous curves) 
are in excellent agreement with
the above predictions \eqref{Etot} and \eqref{tresholdEA}. 
Furthermore we have verified that when $\beta>1/2$ both the independent bit
error approximation and the approximation leading to Eq. \eqref{Etot} fail
as expected. This can be seen for example in Fig. \ref{fig:bgtr}a where we
report the results for the case $\beta=2/3$. Indeed the numerical results
(black dots) differ from the continuous line which corresponds to Eq.
\eqref{Etot}, thus confirming that in this case both the shape of finite size
corrections and the position of the threshold cannot be derived by
\eqref{Etot}.

\subsection{Belief Propagation (BP)}
\label{BP}

The algorithm considered in previous section is very simple since it does not exploit the information contained in the reduced graph (see section \ref{inference}). We will now  instead define a different algorithm in order to extract as much information as we can from the first stage. As already explained in section \ref{inference}, this requires in principle the minimization of Eq. \eqref{infe}.
In order to perform this task we will use Belief Propagation (BP) algorithm to estimate for each variable $i$ the value of the marginal probability $P(x_i)$. Then we will set
to one (to zero) variables for which $P(x_i)>1/2$ (respectively $P(x_i)\leq 1/2$). 
Let us derive the BP equations for the marginal probabilities.
We denote by ${\cal{N}}_i$ (${\cal{N}}_a$) the set of  function (variable) nodes connected to the variable node $i$ (respectively to the function node $a$), by
$P(x_i)^{i\to a}$ the probability of value
$x_i$ for the i-th variable in absence of test $a$
and by $P(x_1,x_2,\dots x_n)^{(a)}$ the joint cavity distribution in the absence of $a$
(so that $P(x_i)^{i\to a}=P(x_i)^{(a)})$).
We can then write
$$P(x_i)^{i\to a}=Ap^{x_i}(1-p)^{1-x_i}\prod_{b\in {\cal{N}}_i\setminus a}
\left(\sum_{\vec x_{\partial_{a,i}}}P(\vec x _{\partial_{a,i}})^{(b)}\id(T_b(x)=t_b)\right)$$
where by $\vec x_{\partial_{a,i}}$ we denote the vector
$\{x_j|j\in {\cal{N}}_a\setminus i\}$. Furthermore we make the usual assumption
that the joint cavity distributions $P(\vec x _{\partial_{a,i}})^{(b)}$ factorize
$$P(\vec x _{\partial_{a,i}})^{(b)}=\prod_{j\in {\cal{N}}_a\setminus i}P^{(b)}(x_j)=
\prod_{j\in {\cal{N}}_a\setminus i}P(x_j)^{j\to b}$$
which leads to closed equations for the set of single variable cavity probabilities.
In order to simplify these equations
we define a normalized message $P(x_i)^{a\to i}$
from function node $a$ to variable node $i$ as
$$P(x_i)^{a\to i}:=C \sum_{j\in {\cal{N}}_a\setminus i}P(x_j)^{(a)}\id(T_a(x)=t_a)
$$ and 
therefore
$$P(x_i)^{i\to a}=Bp^{x_i}(1-p)^{1-x_i}\prod_{b\in {\cal{N}}_i\setminus a}
P(x_i)^{b\to i}$$
and
$$P(x_i)=Bp^{x_i}(1-p)^{1-x_i}\prod_{b\in {\cal{N}}_i}
P(x_i)^{b\to i}.$$
Using the fact that $x_i$ takes values in $\{0,1\}$ and that
both $P^{a\to i}$ and $P^{i\to a}$ are normalized we introduce
cavity fields $h_{i\to a}$ and cavity biases $u_{a\to i}$ defined as follows
$$P(x_i)^{a\to i}=(1-u_{a\to i})\delta_{x_i,0}+u_{a\to i}\delta_{x_i,1}$$
$$P(x_i)^{i\to a}=(1-h_{i\to a})\delta_{x_i,0}+h_{i\to a}\delta_{x_i,1}.$$
The
 BP equation for the cavity biases and fields are:
\begin{numcases}{u_{a\to i}= }
 \nonumber
0 & {\mbox{if}} $t_a=0$\nonumber\\
\left(2-\prod_{j\in {\cal{N}}_a\setminus i} (1-h_{j\to a})\right)^{-1}  & {\mbox{if}} $t_a=1$\nonumber
\end{numcases}
and 
$$h_{i\to a}=\frac{p\prod_{b\in {\cal{N}}_i\setminus a}u_{b\to i}}{p\prod_{b\in {\cal{N}}_i\setminus a}u_{b\to i}+(1-p)\prod_{b\in {\cal{N}}_i\setminus a}(1-u_{b\to i})}.$$
Our detection procedure corresponds to initialize the cavity and bias fields
to some values and iterate BP equations above until they converge. Then, the
marginal probability distribution $P(x_i)$ can be rewritten as
$$P(x_i)=(1-H_{i})\delta_{x_i,0}+H_{i}\delta_{x_i,1}$$
with the full local field $H_i$ satisfying 
$$H_i=\frac{p\prod_{b\in {\cal{N}}_i}u_{b\to i}}{p\prod_{b\in {\cal{N}}_i}u_{b\to i}+(1-p)\prod_{b\in {\cal{N}}_i}(1-u_{b\to i})}$$
and the inference procedure is completed by setting $x_i$ to one (to zero) if
$H_i>1/2$ ($H_i\leq 1/2$ respectively). Note that on the sure variables BP
algorithm lead to the correct detection. Furthermore one should expect that
its performance is better than EA: since we analyze also the information which
comes from tests which are non strippable it is possible that some of the
undetermined ones which are all set to zero in EA are here correctly detected.

\begin{figure}
\begin{center} 
\includegraphics[scale=0.55]{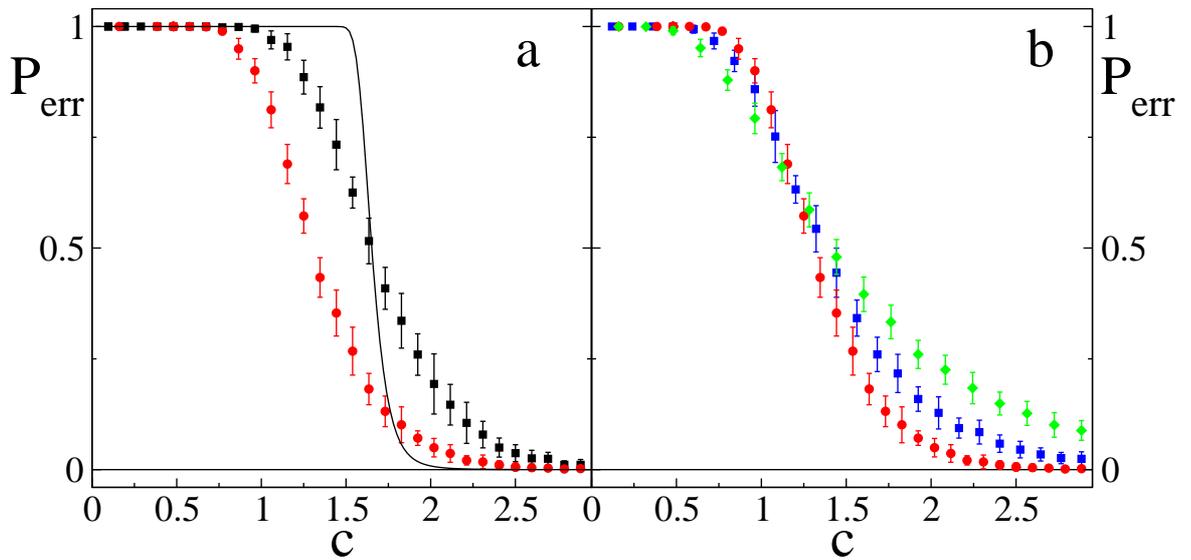}
\end{center}
\caption{{\bf a)} Error probability as a function of $c$ for a regular-regular graph using EA (black squares) and BP (red circles). The graph parameters are chosen as in \eqref{parameters}, with 
$p = N^{-\beta}$, $\beta = 2/3$,  $\alpha=1$ and $N = 2^{15}$. 
The continuous line corresponds to formula \eqref{Etot}. As explained in the text, the discrepancy between the latter and the numerical results confirms that in this regime the approximations leading to \eqref{Etot} are not verified.
{\bf b)} Error probability as a function of $c$ using BP.
We set again $\beta = 2/3$,  $\alpha=1$  and we choose
 $N=2^{15}$ (red circles), $2^{12}$ (blue squares), and $2^{9}$ 
(green diamonds).}
\label{fig:bgtr}
\end{figure}

In order to test the performance of BP algorithm we run the procedure on the
regular-regular graph for $\beta=1/4$ and $\alpha=\log 2$ as we
did for EA. The total error probability as a function of $c$ is reported in
figure \ref{transition}b (black squares). As for EA, a non-detection/detection
phase transition occurs at $\tilde c=1/(\log 2)^2$. Thus, even if EA is a much
simpler algorithm, the performance of the two coincide in the large $N$ limit,
suggesting that for the choice $p=1/N^{\beta}$ with $\beta=1/4$ the reduced
graph does not carry any additional information. In figure \ref{fig:bgtr}a)
we plot instead the total error of EA and BP when $\beta=2/3$ for $N=2^{15}$.
The data indicate that BP algorithm performs much better than the EA in this
case: the reduced graph carries information which is used by BP to optimize
the procedure. We have also verified that the difference between BP and EA
performance does not diminish as the size of the graph is increased. In Fig.
\ref{fig:bgtr}b) we plot instead the results for BP again in the case
$\beta=2/3$ but for different values of $N$. The data become sharper as $N$
is increased. Similarly to the $\beta=1/4$ case, this seems to indicate the
presence of a sharp phase transition in the thermodynamic limit.

Let us start by evaluating the non-detection/detection threshold from BP
equations and then explain why we expect it to coincide with the one for EA at
least when $\beta$ is in $(0,1/3)$. 

If we denote by $\cP^0(H)$ ($\cP^1(H)$) the mean over the random graph
distribution of the probability for the full local field on $i$ conditioned to
the fact that $x_i=0$ ($x_i=1$), the probability of setting to a wrong value
the i-th variable is here $E_{bit}=E_{bit}^0+E_{bit}^1$ with
\begin{equation}
\label{B0}
E_{bit}^0=(1-p)~\int_{\frac{1}{2}}^1 \cP^0(H) dH
\end{equation}

\begin{equation}
\label{B1}
E_{bit}^1=p~\int_0^{\frac{1}{2}} \cP^1(H) dH.
\end{equation}

From the BP equations it is easy to obtain the following 'replica symmetric' cavity equations satisfied by  $\cP^0(H)$ and $\cP^1(H)$
\cite{RichardsonUrbanke}:
\begin{equation}
\cP^{0}(h)=\sum_{l\ge 0}\vdeg_{l} \int \prod_{b=1}^{l} dQ^{0}(u_b)\delta\left(h-\frac{p\prod_b u_b}{p\prod_b u_b+(1-p)\prod_b(1-u_b)}\right)
\end{equation}

\begin{equation}
\cP^{1}(h)=\sum_{l\ge 0}\vdeg_{l} \int \prod_{b=1}^{l} dQ^{1}(u_b)\delta\left(h-\frac{p\prod_b u_b}{p\prod_b u_b+(1-p)\prod_b(1-u_b)}\right)+\evdeg_1 \delta(h-1)
\end{equation}

where

\begin{equation}
\label{ie0}
P^{0}(h)=\sum_{l\ge 1}\evdeg_{l} \int \prod_{b=1}^{l-1} dQ^{0}(u_b)\delta\left(h-\frac{p\prod_b u_b}{p\prod_b u_b+(1-p)\prod_b(1-u_b)}\right)
\end{equation}

\begin{equation}
P^{1}(h)=\sum_{l\ge 1}\evdeg_{l} \int \prod_{b=1}^{l-1} dQ^{1}(u_b)\delta\left(h-\frac{p\prod_b u_b}{p\prod_b u_b+(1-p)\prod_b(1-u_b)}\right)
\end{equation}

\begin{equation}
\begin{split}
Q^{0}(u)=&\sum_k \efdeg_k \int \prod_{j=1}^{k-1} \left[\sum_{y_j} p^{y_j}(1-p)^{(1-y_j)} dP^{y_j}(h_j)\right] \\
&\left[\delta(u)
\prod_{j=1}^{k-1}\delta_{y_j,0}+(1-\prod_j \delta_{y_j,0})\delta\left(u-\frac{1}{2-\prod_j(1-h_j)}\right)\right]
\end{split}
\end{equation}

\begin{equation}
\begin{split}
\label{ief}
Q^{1}(u)=\sum_k \efdeg_k \int \prod_{j=1}^{k-1} &\left[\sum_{y_j} p^{y_j}(1-p)^{(1-y_j)} dP^{y_j}(h_j)\right] \\
&\left[\delta\left(u-\frac{1}{2-\prod_j(1-h_j)}\right)\right]
\end{split}
\end{equation}

It is now easy to verify that $\cP^0(0)=S_0$ and $\cP^1(1)=S_1$, where $S_0$ and $S_1$ are the probability that a variable is sure zero and one respectively, and are given by Eqs. (\ref{S0rr}) and (\ref{S1rr}). Furthermore the following relation holds
$$\cP^0(p)=\cP^1(p)\geq \Lambda[Q^0(1/2)]=\Lambda[Q^1(1/2)]=
\Lambda[1-\rho(1-p \tilde S_1)]=I$$ 
where $I$ is the probability that a variable is isolated, given in Eq. (\ref{Irr}).

By using the above observations together with the definitions \eqref{B0} and \eqref{B1} for the bit error probabilities one obtains
 the following inequalities

\begin{equation}
\label{ine1}
E_{bit}^0\leq (1-p)(1-\cP^0(0)-\cP^0(p))=(1-p)(1-S_0-I)
\end{equation}
\begin{equation}
\label{ine2}
pI=p\cP^1(p)\leq E_{bit}^1\leq p(1-\cP^1(1))=p(1-S_1).
\end{equation}
We will now show how it is possible to locate the non-detection/detection transition from these inequalities without the need to evaluate
the bit error probabilities.

The leading order of the quantities $S_0$, $S_1$ and $I$ have been evaluated in section \ref{poolone}. Furthermore, for $\beta+d<0$ the higher order corrections give $S_0=1-N^d-fN^{-\beta+d}\log N$ and $I=N^d-fN^{d-\beta}\log N$
where $f=\exp(-\alpha)(\alpha/2+1)/(1-\exp(-\alpha))$. Thus
$$N^{-\beta+d}\leq E_{bit}\leq 2f N^{-\beta+d}\log N\ .$$

Therefore, in the assumption of independent bit errors,
we get
$$1-\exp(-N^{1-\beta+d})\leq E_{tot}=1-\left(1-E_{bit}^1-E_{bit}^0\right)^N\leq 1-\exp(-N^{1-\beta+d}\log N)$$
for $\beta+d<0$, namely $c\alpha|\log (1-\exp(-\alpha))|>\beta$. Since $\beta<1/2$ we have $1-\beta>\beta$ and the above bounds on the total error imply the occurrence of a phase transition at the same value $\bar c(\alpha)  $ found with  the EA algorithm (see (\ref{tresholdEA})).
Thus the performance
of EA and BP coincide if the approximations leading to Eqs.
\eqref{S0rr}, \eqref{S1rr} and \eqref{Irr} are correct. By the discussion at
the end of section \ref{poolone} we know that these approximations are under
full control for $\beta<1/6$ and we expect them to hold also up to
$\beta<1/3$. We conclude that in this regime the value of the threshold for BP transition  equals the one for EA
\eqref{tresholdEA}, as is indeed confirmed by the numerical results that we
already discussed for the case $\beta=1/4$ (see Fig. \ref{transition}). We
stress that there is no reason for that to be true in the
regime where the approximations of neglecting proper loops which lead to
\eqref{S0rr}, \eqref{S1rr} and \eqref{Irr} do not hold. For example, as is
shown in Fig.\ref{fig:bgtr}a and b, in the case $\beta=2/3$ even if a sharp
non-detection/detection phase transition seems to occur when $N\to\infty$, the
error probability
is certainly not in agreement with 
\eqref{Etot} which for the chosen parameters would yield to a threshold
at $c\simeq 1.453$.

Note that in the discussion above  
we have upper bounded the bit error with the error over all variables that are
neither sure nor isolated and lower bounded it with the error over isolated
variables. It is thus immediate to see that the position of the phase transition
remains unchanged for all algorithms which set to zero all the isolated
variables (which is the best guess since we have no information and $p<1/2$)
and set to the correct value the sure variables (EA is indeed the simplest
algorithm which belongs to this class). This is due to the fact that
 the mean number of tests in the reduced graph goes to zero in
the detection  regime $-d>{1-\beta}>2/3$, as can be checked using formula \eqref{reducible} and neglecting loops.

Finally, we would like to stress that even if we have shown that EA and BP
inference procedures are optimal for R-R and P-R pool designs, at least when
$\beta<1/3$, 
this does not imply that these pool designs are optimal over all the possible designs of the factor graph.
However, an indication that they might be optimal comes from the results
on two-stage exact algorithms presented in section \ref{two}. As a further
check we have evaluated the thresholds for the Poisson-Poisson (P-P) and
Poisson-regular (P-R) cases. 
Using the same technique as above, we found in both cases a 
 non-detection/detection phase transition which occurs at the same threshold for EA and BP.
If we set $K=\alpha/p$, $M=c\alpha\log N$, $L=c\alpha\log p$ the threshold value is
\begin{equation}
\label{tresholdPP}
\bar c(\alpha)=\frac{1-\beta}{\alpha\exp(-\alpha)}.
\end{equation}
By optimizing \eqref{tresholdPP} over the choice of $\alpha$ we get $\tilde \alpha=1$ and $\overline M=eNp|\log p|$,
which is  larger than the optimal threshold for R-R and R-P.

\section{Two-stage algorithms}
\label{two}

In this section we analyze two-stage exact algorithms when the number of
items, $N$, goes to infinity and the defect probability, $p$, goes to zero as
$p=1/N^{\beta}$. This setting was first discussed by Berger and Levenshtein in
\cite{Berger} where they proved that if $0<\beta<1$, the minimal (over all
two-stage exact procedures) mean number of tests, $\overline T(N,p)$,
satisfies the bounds
$$\frac{1}{\log 2}\leq \lim_{N\to\infty} \frac{\overline T(N,p)}{Np|\log p|}\leq \frac{4}{\beta}.$$
In \cite{MezTon} two of the authors have derived the prefactor for the above scaling when
$0\leq
\beta<1/2$,
 \begin{equation}\label{exact}\lim_{N\to\infty}\frac{\overline
     T(N,p)}{Np|\log p|}=\frac{1}{(\log 2)^2}\ 
 \end{equation}
 and constructed a
 choice of algorithms over which this optimal value is attained. Note that our
 analysis includes the case $\beta=0$, namely the situation in which the limit
 $p\to 0$ is taken after $N\to\infty$. Note that the asymptotic result
 \eqref{exact} is $1/\log 2$ above the information theoretic bound 
 $\overline T(N,p)\geq Np|\log p|/\log 2$. In section \ref{twoone} we give a
 short account of the derivation of \eqref{exact} and we construct an optimal
 algorithm. In section \ref{twotwo} we test the performance of algorithms
 corresponding to different choices of the random pools of the first stage.

\subsection{Optimal number of tests for $p=1/N^{\beta}$, $~~\beta\in(0,1/2]$}
\label{twoone}

An exact two-stage algorithm involves a first stage of tests
after which all variables are identified and set to their value. Then a second stage is performed where all the remaining variables are individually tested. 
The mean number of tests, $T(N,p)$, is therefore given by 
\begin{equation}
\label{tests}
T(N,p)=M+N-\sum_{i=1}^N (p^i_{s0}+p^i_{s1})
\end{equation}
where $M$  is the number of tests of the first stage and $p^i_{s0}$ and $p^i_{s1}$ are the probabilities for variable $i$ to be sure zero and sure one. The latter in turn are given by Eqs. \eqref{ps0} and \eqref{ps1} with ${\cal{N}}_a$'s and ${\cal{N}}_i$'s being the neighborhood of tests and variables  of  the first stage. 

It is immediate to verify that in the limit $N\to \infty$ and $p\to 0$ the number of individual check over undetected ones is irrelevant, i.e.
\begin{equation}
\label{ine}
\frac{T(N,p)}{Np|\log p|}=\frac{M+N-\sum_{i=1}^Np^i_{s0}}{Np|\log p|}
\end{equation}
Furthermore $p^i_{s0}$ is always upper bounded by the expression 
\eqref{ps0noloops} obtained by neglecting loops, as is proven in \cite{MezTon} by using Fortuin-Kasteleyn-Ginibre inequality \cite{FKG} together with the observation that the existence of at least
one variable equal to one in two (or more) intersecting pools are positively correlated. We define $f(\vec m)$ to be the fraction of sites such that among their neighbors there are $m_1$ tests of degree $1$, $m_2$ tests of degree $2$, etc. By using \eqref{ps0noloops} and \eqref{ine}, the optimal number of tests over all two stage procedures can be lower bounded as
\begin{equation}
\label{ine2}
\frac{\overline T(N,p)}{Np|\log p|}\geq \inf_{f(\vec m)}\left(\frac{\sum_{\vec m}f(\vec m)\left(\sum_{j=1}^N\frac{m_j}{j}+(1-p)P(\vec m)\right)}{p|\log p|}\right)
\end{equation}
where the infimum is over all possible probability distributions $f:(1,\dots N)^N\to \mathcal{R}^+$ with $\sum_{\vec m}f(\vec m)=1$ and
\begin{equation}
P(\vec m)=\prod_{i=1}^N(1-(1-p)^{j-1})^{m_j}.
\end{equation}
Minimization over $f(\vec m)$ can then be carried out and leads in the limit $p\to 0$ to
\begin{equation}
\label{ine3}
\frac{\overline T(N,p)}{Np|\log p|}\geq \frac{1}{(\log 2)^2}.
\end{equation}
Furthermore the above minimization procedure shows
that this infimum is attained for 
$f(\vec m)=\delta_{\vec m,\bar m}$ with $\bar m_i=\delta_{i,\log 2/p}[|\log p|/\log 2]$. This implies that the lower bound is saturated on the uniform distribution over regular-regular graphs
with $L=[|\log p|/\log 2]$ and $K=[\log 2/p]$ provided that we can neglect loops in the evaluation of $p^i_{s0}$. This, as already explained in section \ref{poolone}, is true as long as $\beta<1/2$. Note  that the optimal result is also attained if instead of a random construction of pools we fix a regular-regular graph  which has no loops of length $4$ and has  the same choices of test and variable degrees as above. The existence of at least one of such a graph for these choices of $K$ and $L$ when $\beta<1/2$ is guaranteed by the results in \cite{LuMoura}. Thus we have established the result \eqref{exact} for the optimal value of tests over all exact two-stage procedure and constructed algorithms based on regular-regular graphs which attain this optimal value.

\subsection{Testing different pool designs for $p\to 0$}
\label{twotwo}

We will now check the performance of different pool designs corresponding to different random distributions for the pools in the first stage. In all cases we will  fix the degree profiles $\Lambda$ and $P$ and consider a uniform distribution over graphs with these profiles. Using the notation of section \ref{inference} and neglecting the presence of loops, the mean number of tests \eqref{tests}
can easily be rewritten 
\begin{equation}
\begin{split}
\frac{T(N,p)}{N}=&\frac{\sum_k \efdeg_k/k}{\sum_l \evdeg_l/l}+ (1-p)\vdeg\big[1-\efdeg[1-p]\big]\\
 &+ p \vdeg\Big[1-\efdeg\big[(1-p)(1-\evdeg[1-\efdeg[1-p]])\big]\Big]
\end{split}
\label{bigResult}
\end{equation}
(we suppose that  the fraction of both test and variable nodes of degree zero
is equal to zero).
As for the one stage case, we consider four different choices of the connectivity distributions corresponding to regular-regular (R-R), regular-Poisson (R-P), Poisson-Poisson (P-P) and Poisson-regular (P-R) graphs and for each choice we have optimized over the parameters of the distribution. The corresponding degree profiles and edge perspectives are given in section \ref{EA} and \ref{BP}. The first term of the r.h.s. of Eq. (\ref{bigResult}) corresponds to the total number of tests of the first stage per variable, i.e. $L/K$, while the second and third terms correspond to $(1-p)(1-S_0)$ and $pS_1$ respectively, where $S_0$ and $S_1$ have already been evaluated in the previous section
(see Eqs. \eqref{S0rr}, \eqref{S1rr}, \eqref{S0rp}, \eqref{S1rp}).

We now  let $K=\alpha/p$ and $L=c\alpha|\log p|+v$ (in order to keep corrections in $M$ to the leading term $Np|\log p|$) and we evaluate \eqref{bigResult} for the different pool designs. Then we optimize over the parameters $\alpha$ and $c$.

\subsubsection{Regular-Regular and Regular-Poisson case}

If we set $d=c\alpha|\log(1-\exp(-\alpha))|$, both in the R-R and  R-P case we get
\begin{equation}
\begin{split}
\frac{T(N,p)}{N}=cp|\log p|+vp/\alpha+p^d(1-\exp(-\alpha))^v+o(p^{1+d}).
\end{split}
\label{RR}
\end{equation}
Thus the optimal value for $p\to 0$ is given by $d=1$, namely $$c(\alpha)=\frac{1}{\alpha|\log(1-\exp(-\alpha))|}.$$
By optimizing over $\alpha$ we get $\bar\alpha=\log 2$ and $\bar c=1/(\log 2)^2$.
Then minimizing over $v$ we get
\begin{equation}
\frac{T}{Np}= \left(\frac{1}{\log 2}\right)^2 \left(|\log p|+1+2 \log\log2 \right)
\end{equation}
\subsubsection{Poisson-Poisson and Poisson-Regular case}

If we set $f=c\alpha\exp(-\alpha)$, for both the P-P and P-R case we get
\begin{equation}
\begin{split}
\frac{T(N,p)}{N}=cp|\log p|+vp/\alpha+p^f\exp(-v\exp(-\alpha))+o(p^{1+f}).
\end{split}
\label{PP}
\end{equation}
Thus the optimal value for $p\to 0$ is given by $f=1$, namely $$c(\alpha)=\frac{1}{\alpha\exp(-\alpha)}.$$
By optimizing over $\alpha$ we get $\bar\alpha=1$ and $\bar c=e$.
Then minimizing over $v$ we get $v=-e$, thus
\begin{equation}
\frac{T}{Np}= e|\log p|+o(p^f).
\end{equation}



\subsection{Optimal algorithms at finite $p$}

\begin{figure}
\begin{center} 
\psfrag{uffa}[][]{{\Huge{$T/NH(p)$}}}
\psfrag{ciao}[][]{{\Huge{$\log(p)$}}}
\includegraphics[angle=270,scale=0.4]{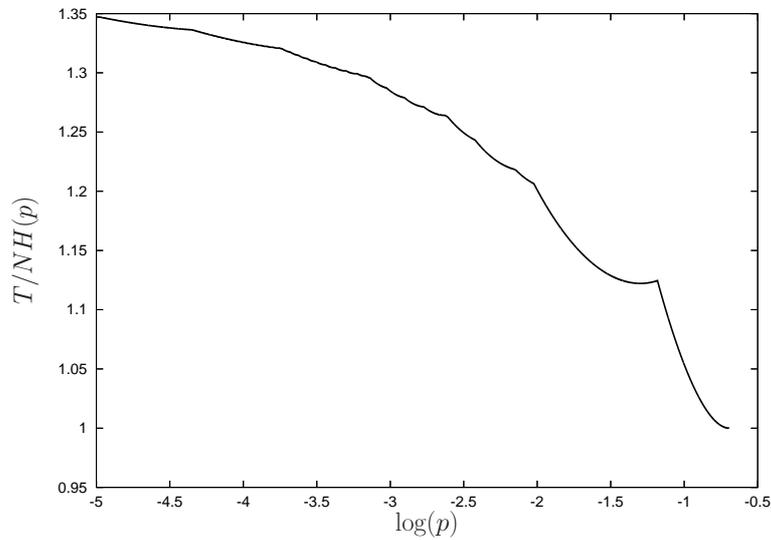}
\end{center}
\caption{Expected mean number of tests divided by the information theoretic lower bound $NH(p)=N(p\log_2 p+(1-p)\log_2(1-p))$ for the regular-regular graphs which optimize \eqref{bigResult}. The non-analyticity points correspond to the values of $p$ where the optimal degree pair $L,K$ changes, see Fig.\ref{pfinite2}. In the small $p$ limit the curve goes asymptotically to $1/\log 2$ in agreement with \eqref{exact}.}
\label{pfinite1}
\end{figure}

\begin{figure}
\begin{center} 
\psfrag{uffa}[][]{{\Huge{$L,~~ \log K$}}}
\psfrag{ciao}[][]{{\Huge{$\log(p)$}}}
\includegraphics[angle=270,scale=0.4]{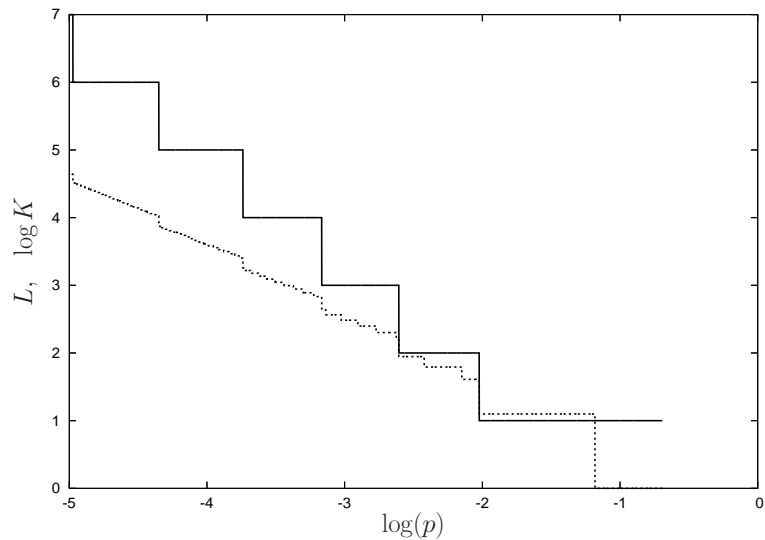}
\end{center}
\caption{Values of $L$ (continuous line) and of $\log K$ (dotted line) corresponding to the couples $L,K$ which give  the optimal mean number of tests of Fig.\ref{pfinite1}}
\label{pfinite2}
\end{figure}

The above results show that both for regular-regular and regular-Poisson
graphs the optimal asymptotic value \eqref{exact} can be reached in the case
$p\to 0$, while this is true neither in the Poisson-Poisson nor in the
Poisson-regular case. Note however that this does not exclude the existence of
other distributions for which the optimal value is attained. We stress once
more that even if when we performed optimization we did not make any
assumption on how $p\to 0$, the results hold only if proper loops can be
neglected in the resulting optimal graphs. This includes the following
regimes:  either $p\to 0$ after $N\to\infty$ or  $p=1/N^{\beta}$ with
$\beta<1/2$. The reason why we focused on the $p\to 0$ limit is twofold. On
the one hand one often deals in practical applications with problems in which
the defective probability is small. On the other hand the information
theoretic lower bound $T(N,p)\geq Np|\log p|/\log 2$ already tells us that if
$p\not\to 0$ the number of tests is proportional to $N$ as  in the
trivial procedure which tests all variables individually. However one could be interested in the optimal random pool design for the first stage if
instead $p$ is held fixed. A natural conjecture in view of the results of the
previous sections is that, at least for sufficiently small $p$, this
corresponds again to a regular-regular graph. In order to solve this problem
one should find the best degree sequences $\Lambda,P$ which minimize the
expression \eqref{bigResult}. This is a hard minimization problem which we
simplified by first proving that (for a general choice of $N$ and $p$) at most
$3$ coefficients $\Lambda_l$ and at most $5$ coefficient $P_r$ are non zero in
the optimal sequence. Plugging this information in some numerical minimization
procedure of (\ref{bigResult}), we have observed that for most values of $p$
the optimal degree sequence is the regular-regular one. There are
also some values where the optimal graph is slightly more complicated. For
instance for $p=.03$, the best sequences we found are $\Lambda[x]=x^4$ and
$P[x]=.45164\; x^{21}+.54836\; x^{22}$, giving
$T=.25450$
, slightly better than the one obtained 
with the optimal regular-regular one,
$\Lambda[x]=x^4$ and $P[x]=x^{22}$, giving 
$T=.25454$
. But for all
values of $p$ we have explored, we have always found that either the
regular-regular graph is optimal, or the optimal graph has superposition of
two neighboring degrees of the variables, as in this $p=.03$ case.
In any case regular-regular is always very close to the optimal structure.
In Fig. \ref{pfinite1} we depict the expected mean number of tests (divided by the information theoretic lower bound $NH(p)=N(p\log_2 p+(1-p)\log_2(1-p))$) obtained by the numerical minimization of \eqref{bigResult} on the ensemble of regular-regular graphs.  In the small $p$ limit the curve goes asymptotically to $1/\log 2$ as predicted by \eqref{exact}. In Fig.\ref{pfinite2} we depict instead the corresponding optimal degree couples $K,L$. Note that the non-analyticity points for the expected mean number of tests correspond to the values of $p$ where the optimal  degree pair $L,K$ changes.

\section{Perspectives} \label{conclusions}

As recalled in the introduction, Group Testing strategies are used in a variety of situations ranging from molecular biology to computer science \cite{Dorfman}--\cite{review1}. In most of the applications it is important to take into account the possibility of errors in the tests answers  \cite{Zenios,Rish,Gupta,Macula,Knillerr}, i.e. to consider the {\sl faulty-case} instead of the {\sl gold-standard case} analyzed in this work.
BP equations for cavity biases and fields analogous to those of Section \ref{BP} can be derived also
in the faulty setting and a natural development of the present work is to analyze the performance of the corresponding BP algorithm. A similar task has been performed in \cite{Rish} for a setting relevant for fault diagnosis in computer networks. 

It is important to notice that the relevant form of the test errors  depends on the specific application at hand. 
In the majority of the situations in which GT is a useful tool, one can assume that the errors occur independently in different pools. Thus the error model is completely defined by the probability of {\sl false positive} and {\sl false negative} answers, which are usually either pool independent or they depend only on the size of the pool. An example of the latter situation is given by blood screening experiments for which the false negative probability  increases with the size of the pools due to the inevitable dilution effect \cite{Zenios,Gupta}.

Finally, it is important to bear in mind that, at variance with our analysis, in practical situations one should take into account finite size corrections  as well as the fact that the maximal size of the pool may be limited by experimental constraints.

\vspace{1 cm} 

{\bf {\sl Acknowledgments:}}
This work has been supported in part by the EC grant ``Evergrow'', IP 1935 of FET-IST.

\end{document}